\documentclass[a4paper,reqno,superscriptaddress,twocolumn,showkeys,showpacs]{revtex4}

\usepackage[centertags]{amsmath}
\usepackage{amsfonts}
\usepackage{amssymb}
\usepackage{amsthm}
\usepackage{newlfont}
\usepackage{stmaryrd}
\usepackage{mathrsfs}
\usepackage{euscript}
\usepackage{graphicx}
\usepackage{multirow}

\usepackage{wrapfig}
\usepackage{subfigure}
\usepackage{amscd}

\usepackage{color}

\usepackage[finalnew]{trackchanges}
\addeditor{JD}

\theoremstyle{plain}

\theoremstyle{definition}

\theoremstyle{remark}
  
 \let\be=\beta \let\de=\delta 
\let\ve=\varepsilon   
 \let\la=\lambda \let\om=\omega 
   
\let\ze=\zeta 

\let\De=\Delta

\newcommand{\opunit}{\text{1}\kern-0.22em\text{l}}

\newcommand{\frq}{{\mathfrak q}}

\DeclareMathAlphabet{\mathpzc}{OT1}{pzc}{m}{it}

\newcommand{\cf}{cf.\;}

\newcommand{\ie}{i.e.\;}

\newcommand{\rel}{\,|\,}
\newcommand{\id}{\textrm{d}}

\newcommand{\sgn}{\operatorname{sgn}}

\begin{document}

\title{Model study on steady heat capacity in driven stochastic systems}

\author{Ji\v{r}\'i Pe\v{s}ek}
\affiliation{Institute of Physics,  Academy of Sciences of the Czech Republic, 18221 Prague, Czech Republic}
\email{pesek@fzu.cz}
\affiliation{Faculty of Mathematics and Physics, Charles University, 12116 Prague, Czech Republic}
\author{Eliran Boksenbojm}
\affiliation{Instituut voor Theoretische Fysica, K.U.Leuven, B-3001 Leuven, Belgium}
\author{Karel Neto\v{c}n\'{y}}
\affiliation{Institute of Physics,  Academy of Sciences of the Czech Republic, 18221 Prague, Czech Republic}
-


\begin{abstract}
We explore two- and three-state Markov models driven out of thermal equilibrium by non-potential forces\add[JD]{,} to demonstrate basic properties of the steady heat capacity based on the concept of quasistatic excess heat. It is shown that large enough driving forces can make the steady heat capacity negative. For both the low- and  high-temperature regimes we propose an approximative thermodynamic scheme in terms of ``dynamically renormalized'' effective energy levels.
\end{abstract}


\maketitle

\note[JD]{General comments from language editor - english is good and the paper reads well overall.  A number of minor alternations are suggested following a more detailed review.}

\section{Introduction}

Statistical mechanics away from thermal equilibrium and detailed balance is
an important field of modern theoretical physics with many open problems, one of them being a systematic characterization of heat processes connecting nonequilibrium steady states. In 
this paper
we focus on one specific step \change[JD]{within this program}{therein}, namely \remove[JD]{on} the problem \add[JD]{of} how to naturally extend the concept of heat capacity to steady nonequilibrium systems. A specific proposal motivated by \remove[JD]{some} previous work on macroscopic frequency-dependent calorimetry~\cite{sta,cera,nd}\add[JD]{,} and reformulated within the framework of nonequilibrium stochastic processes\add[JD]{,} has been given in~\cite{new}\remove[JD]{,} based on the concept of quasistatic excess heat~\cite{pump,ha}.

As it has been observed in~\cite{new}, the steady heat capacity exhibits some new features with no direct equilibrium analogy, including 
the possibility that it becomes negative
far from thermal equilibrium. In this paper we continue the study by analyzing two simple discrete models which can be considered as paradigms: (1) a stochastic two-level model with multiple transition channels and (2) a three-level model\change[JD]{, b}{. B}oth \add[JD]{are} maintained under nonequilibrium conditions via the presence of non-potential forces. We analyze their thermal properties in various temperature and driving regions\add[JD]{,} and we propose \change[JD]{some}{an} interpretation of the obtained results in terms of ``renormalized'' (thermo)dynamic quantities. We pay \remove{a} special attention to a ``gauge'' invariance that is inherently present in systems with non-potential forces and we \change[JD]{will}{also} indicate how it can be used in a constructive way.

The paper is organized as follows. In the next section we review the formalism of discrete Markovian processes that can describe nonequilibrium systems and we discuss their gauge invariance. We also introduce the two nonequilibrium models \add[JD]{mentioned above,} and analyze their basic steady properties. In section~\ref{sec:heatcapacity} we explain the concept of excess heat in the quasistatic limit. Our main results for the two models are given in Section~\ref{sec:results} and we conclude with a summary and outlook towards further research work, section~\ref{sec:conclusion}.

\section{Model nonequilibrium systems}\label{sec:models}

We consider stochastic systems with discrete states representing distinct configurations of some 
\annote[JD]{mesoscopic}{do you mean microscopic or macroscopic?} 
system that exchanges energy with its surrounding\add[JD]{s} in \change[JD]{terms}{the forms} of work and dissipated heat. As a slight generalization with respect to~\cite{new}, we allow for multiple transition channels $i \stackrel{b}{\leftrightarrow} j$ connecting any two states $i$ and $j$ along a channel $b$.
Each transition channel is associated with the amount of heat
$q_{i \rightarrow j}^b = -q_{j \rightarrow i}^b$
dissipated into an attached reservoir. Typically, the heat quanta are introduced via the energy balance along the transitions\add[JD]{,} with some \emph{a priori} given energy levels $\ve_i$ of the states together with the amount of work
$w_{i \rightarrow j}^b$ done on the system by other forces whose effect is not included in the energy level differences. The energy balance along any single transition
$i \stackrel{b}{\rightarrow} j$ is
\begin{equation}\label{balance-micro}
  \ve_j - \ve_i = w_{i \rightarrow j}^b - q_{i \rightarrow j}^b
\end{equation}
A general stochastic trajectory $\om$ consists of a sequence of such transitions\add[JD]{,} and the global balance equation then reads
$\ve_\text{fin} - \ve_\text{ini} = \sum^{(\om)} w_{i \rightarrow j}^b -
\sum^{(\om)} q_{i \rightarrow j}^b$, with $\sum^{(\om)}$ running over all transitions
along $\om$. Here we assume the system's environment to consist of a \emph{single} heat bath so that the total dissipated heat
$Q(\om) = \sum^{(\om)} q_{i \rightarrow j}^b$ equals the energy increase in the bath which can \emph{in principle} be measured by calorimetric techniques.

We assume here a typical scenario of nonequilibrium stochastic thermodynamics in which the system is \emph{driven} out of equilibrium by the presence of non-conservative forces~\cite{seifert}. Since such \remove[JD]{a} force\change[JD]{ does}{s do} not derive from a potential, \change[JD]{its}{their} work \emph{cannot} be completely included in the energy level differences and
$W(\om) = \sum^{(\om)} w_{i \rightarrow j}^b$ depends on the entire trajectory $\om$ rather than on the initial and final states only.
In this way the energy balance equation~\eqref{balance-micro} gives the heat in terms of the work of potential and non-potential forces, respectively. In applications there is often a natural way \remove[JD]{how} to make such a decomposition so that both potential and non-potential components can be accessed separately. Yet, this decomposition is intrinsically non-unique: the ``gauge'' transformation
\begin{equation}\label{gauge}
\begin{split}
  \ve_i &\mapsto \hat\ve_i = \ve_i + \psi_i\,,
\\
  w_{i \rightarrow j}^b &\mapsto
  \hat w_{i \rightarrow j}^b = w_{i \rightarrow j}^b + \psi_j - \psi_i
\end{split}
\end{equation}
with an arbitrary $\psi$ leaves the dissipation functions
$q_{i \rightarrow j}^b$ invariant\add[JD]{,} and 
this
can sometimes be used to simplify the description of a model. The simplest example is a close-to-equilibrium regime with the gauge fixed so that the non-potential forces (or the corresponding work functions) become manifestly small, and \change[JD]{hence they can}{can hence} be dealt with as \remove[JD]{a} perturbation\add[JD]{s} around a reference equilibrium.

Later we will show how to employ the gauge invariance even in some far-from-equilibrium regimes. The idea is to 
adapt
the gauge to a certain dynamical potential that naturally appears in the context of quasistatic energy exchange, and which itself is gauge-invariant though generally temperature-dependent. Loosely speaking, those regimes where the temperature-dependence  becomes weak allow \change[JD]{to construct}{for the construction of} ``almost well-defined'' energy levels.\\

We assume the dynamics of our models to be Markovian, with
$k_{i \rightarrow j}^b$ the transition rates that are related to the heat functions
$q_{i \rightarrow j}^b$ by the \emph{local detailed balance principle}~\cite{LS}:
\begin{equation}\label{local-db}
  k_{j \rightarrow i}^b = k_{i \rightarrow j}^b \exp \Bigl(
  -\frac{q_{i \rightarrow j}^b}{k_B T} \Bigr)
\end{equation}
where $T$ is the reservoir\remove[JD]{'s} temperature.
Obviously, \change[JD]{on}{in} the absence of non-potential forces (or, in terms of the gauge transformation~\eqref{gauge}, if the work contribution to the balance~\eqref{balance-micro} can be completely transformed out \change[JD]{to have}{such that} $\hat w_{i \rightarrow j}^b \equiv 0$), the above condition boils down to the usual detailed balance,
$k_{j \rightarrow i}^b = k_{i \rightarrow j}^b \exp
[(\hat\ve_i - \hat\ve_j) / (k_B T)]$. This \change[JD]{is an}{equation describes} equilibrium dynamics with Boltzmann steady distribution.
However, the presence of non-potential forces makes the dynamical properties fundamentally different from thermal equilibrium.
The steady state is generally specified by the stationary occupations $\rho_i$ and the stationary currents
$J_{i \rightarrow j}^b = \rho_i k_{i \rightarrow j}^b - \rho_j k_{j \rightarrow i}^b$, obtained by solving the stationary master equation
\begin{equation}\label{stationary}
  \sum_{j,\,b} J_{i \rightarrow j}^b = 0
\end{equation}
with the sum over all target configurations $j$ and all transition channels $b$ connecting both ends. The amount of steady dissipation is given by the total stationary heat current into the reservoir, \ie from all transition channels added together:
\begin{equation}\label{dissipation}
\begin{split}
  \frq &= \frac{1}{2}\sum_{i,\,j,\,b} q_{i \rightarrow j}^b J_{i \rightarrow j}^b
\\
  &= \frac{1}{2}\sum_{i,\,j,\,b} w_{i \rightarrow j}^b J_{i \rightarrow j}^b
\end{split}
\end{equation}
where the second expression follows by combining with the balance~\eqref{balance-micro} and the stationary equations~\eqref{stationary}; the factor $1/2$ being added to avoid double counting.

\subsection{Model I: Two-level system with biased channels}\label{sec:2-level}

\begin{figure}
  \includegraphics[width=5cm]{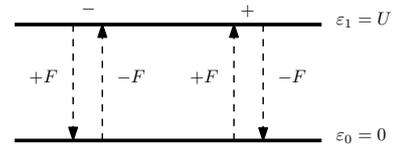}\\
  \caption{Two-level model with two transition channels. The work performed along the counterclockwise cycle equals 2F.}\label{2level}
\end{figure}

As our first model we consider a system with two states `$0$' and `$1$' that are connected by two distinct channels `$+$' and `$-$', see Fig.~\ref{2level}.
We assume that the states have the energies
$\ve_0 = 0$ and $\ve_1 = U > 0$, and that an additional driving force performing work
$w_{0 \rightarrow 1}^\pm = \pm F$ is applied. Hence, for the loop \change[JD]{made of}{formed by} the allowed transitions we have
$W(0 \stackrel{+}{\rightarrow} 1 \stackrel{-}{\rightarrow} 0) = 2F$, manifesting a non-potential character of the driving force. From the heat functions
$q_{0 \rightarrow 1}^\pm = \pm F - U$ we immediately see that the case
$F > U$ (respectively $-F > U$) corresponds to a strong nonequilibrium regime in which the system dissipates a positive amount of energy along both transitions in the loop
$0 \stackrel{+}{\rightarrow} 1 \stackrel{-}{\rightarrow} 0$ (respectively its reversal). Note that in this regime the (original) notion of energy gap separating both states and uniquely distinguishing between the ground and excited states becomes essentially meaningless.

The most general transition rates compatible with the local detailed balance principle~\eqref{local-db} are
\begin{equation}
\begin{split}
  k_{0 \rightarrow 1}^\pm &= A\,\exp \Bigl( \frac{\pm\Phi}{2} +
  \frac{\pm F - U}{2 k_B T} \Bigr)
\\
  k_{1 \rightarrow 0}^\pm &= A\,\exp \Bigl( \frac{\pm\Phi}{2} -
  \frac{\pm F - U}{2 k_B T} \Bigr)
\end{split}
\end{equation}
with arbitrary parameters $A$ and $\Phi$, possibly dependent on the other system parameters $F$, $U$ and $T$ in a non-trivial way. While $A$ only sets an overall  time-scale and hence can be mostly ignored, the parameter $\Phi$ measures the  direction-independent asymmetry between both channels, i.e., their \emph{relative} time-scales. In the sequel we only assume that $\Phi$ can be kept constant when changing the temperature. This appears to be a rather serious (and physically not well motivated) assumption\add[JD]{,} but we make it in order to simplify our presentation as well as to better separate the non-potential and the additional channel-asymmetry effects. For convenience, we set $A = 1$, $k_B = 1$ and always assume $\Phi \geq 0$. (Note the symmetry of the dynamics under the channel-reversal combined with
$F \mapsto -F$ and $\Phi \mapsto -\Phi$.)

The stationary distribution satisfying~\eqref{stationary} coincides with that of the channel-unresolved two-level system having the rates
$\la_{0 \rightarrow 1} = k_{0 \rightarrow 1}^+ + k_{0 \rightarrow 1}^-$ and
$\la_{1 \rightarrow 0} = k_{1 \rightarrow 0}^+ + k_{1 \rightarrow 0}^-$, hence the relative occupations are
\begin{equation}\label{2-occupation}
  \frac{\rho_1}{\rho_0} = \frac{\la_{0 \rightarrow 1}}{\la_{1 \rightarrow 0}} =
  e^{-\frac{U}{T}} \frac{1+\ze}{1-\ze}
\end{equation}
with $\ze = \tanh(\frac{\Phi}{2}) \tanh(\frac{F}{2T})$. The corresponding steady rate of dissipation~\eqref{dissipation} is $\frq = 2 F J \geq 0$ with
$J = J_{0 \rightarrow 1}^+ = J_{1 \rightarrow 0}^-$ the stationary current,
\begin{equation}
  J = \frac{\sinh\bigl(\frac{F}{2T}\bigr)}{\cosh\bigl(\frac{\Phi}{2}\bigr) \cosh\bigl(\frac{U}{2T}\bigr) \bigl[ 1 - \ze \tanh\bigl(\frac{U}{2T}\bigr) \bigr]}
\end{equation}

In formula~\eqref{2-occupation} we see modifications with respect to the equilibrium Boltzmann statistics whenever $\Phi \neq 0$. It agrees with our intuition that \remove[JD]{the} relative throttling of the `$-$' with respect to the `$+$' channel under a positive $F > 0$ tends to increase the occupancy of the ``excited'' state `1'. Eventually in the limit $\Phi \to +\infty$ the `$-$' channel completely closes and the system is again found at thermal equilibrium ($J = 0$) but now with the energy gap $U - F$. The resulting population inversion for $F > U$ is a most simple example of gauge transformation, applied here to easily deal with the driving forces when they become derivable from a potential. From this point of view, the population inversion is a superficial concept here since $U - F = \hat\ve_1 - \hat\ve_0$ is just the full energy gap after the transformation~\eqref{gauge} with $\psi_1 - \psi_0 = -F$ completely removing the driving force has been applied.

However, more important is how these simple observations carry over when both channels remain open to hold the system out of equilibrium: \change[JD]{O}{o}ne checks that for an arbitrarily weak channel asymmetry the population inversion
$\rho_1 > \rho_0$ still occurs whenever the driving force is strong enough\add[JD]{,} and for large enough (but finite) temperatures. This follows from
\begin{equation}\label{2level-rho-highT}
  \log\Bigl(\frac{\rho_1}{\rho_0}\Bigr) =
  -\frac{U - F \tanh\bigl(\frac{\Phi}{2}\bigr)}{T} +
  O\Bigl(\frac{1}{T^2}\Bigr)
\end{equation}
In contrast to the limiting 
case $\Phi \to +\infty$, 
the driving force now does \emph{not} derive from a potential and hence cannot be transformed out.
Nevertheless, the leading term in the high-temperature expansion~\eqref{2level-rho-highT} suggests that
$U - F \tanh\bigl(\frac{\Phi}{2}\bigr)$ may take over the role of an effective energy gap, though we are now dealing with a genuine nonequilibrium system where the energy levels are ambiguously defined. This proposal will be discussed in Section~\ref{sec:results}.

In the low-temperature regime the relative occupation~\eqref{2-occupation} has the asymptotics
$\log(\rho_1 / \rho_0) = -U / T + \Phi \sgn(F) + O(T)$, showing that 
independently of 
the driving force there is no population inversion at zero temperature. Nevertheless, the system undertakes a transition between ``insulator'' and ``conductive'' regimes at
$F = \pm U$ as seen from the low-temperature current asymptotics,
\begin{equation}
  J \simeq \sgn(F)\,e^{\frac{|F| - U}{2T} + \frac{\phi}{2} \sgn(F)}
  \stackrel{T \to 0+}{\longrightarrow}
  \begin{cases}
    0 & \text{if } |F| < U
  \\
    \pm\infty & \text{if } \pm F > U
  \end{cases}
\end{equation}
This can be understood by observing that in the low-driving (or insulator) regime,
$|F| < U$, the state `$0$' remains a well defined ground state in the sense
that in both channels $\log(k_\downarrow^\pm / k_\uparrow^\pm) \to +\infty$ for
$T \to 0$, whereas in the high-driving (or conductive) regime $F > U$ the system exhibits a limit cycle behavior,
$\log(k_\downarrow^- / k_\uparrow^-) \to +\infty$ and
$\log(k_\downarrow^+ / k_\uparrow^+) \to -\infty$; analogously for $-F > U$.

Models with more states can exhibit \change[JD]{still a}{an even} richer behavior as we shortly demonstrate via our second example.

\subsection{Model II: Driven three-level system}
\label{sec:3-levels-intro}

\begin{figure}
  \includegraphics[width=5cm]{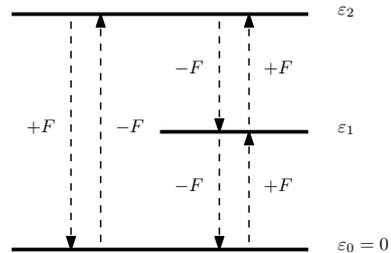}\\
  \caption{Nonequilibrium three-level model.}\label{3level}
\end{figure}
Now we consider a three-level version of the above model, with states `$0$', `$1$' and `$2$' mutually connected by single channels. The system is driven out of equilibrium by a force acting along the loop $0 \rightarrow 1 \rightarrow 2 \rightarrow 0$ and performing equal work,
$w_{0 \rightarrow 1} = w_{1 \rightarrow 2} = w_{2 \rightarrow 0} = F$, along all those transitions, see Fig.~\ref{3level}. The dynamics \change[JD]{is}{are} defined by the transition rates
\begin{equation}
 k_{i \rightarrow i\pm} =
 A_{i,i\pm}\,\exp \Bigl( \frac{\ve_i - \ve_{i\pm} \pm F}{2 k_B T} \Bigr)
\end{equation}
where $i+$ (respectively $i-$) is the succeeding (respectively the preceding) state along the oriented loop $0 \rightarrow 1 \rightarrow 2 \rightarrow 0$; e.g., $0+ = 1$, $0- = 2$ etc.
The prefactors $A_{i j} = A_{j i} > 0$ are symmetric \add[JD]{in order} to satisfy the local detailed balance condition~\eqref{local-db} but arbitrary otherwise.
We only assume \add[JD]{that} they can be kept constant \change[JD]{independently}{and independent} of other parameters like the temperature or forces.
To be specific, we assume that $\ve_2 > \ve_1 > \ve_0 = 0$ and $F > 0$.

The present model exhibits a rich collection of different zero-temperature phases which are summarized in Table~\ref{table_cases}.
In particular, it demonstrates a population inversion
between levels `$0$' and `$1$' in the case $\ve_2 > 2\ve_1$ and $F > \ve_1$. Later we will see that \emph{at} the ``critical'' driving $F = \ve_1$\add[JD]{,} where the zero-temperature population inversion occurs, the model exhibits an anomalous low-temperature behavior.
\begin{table}
\begin{tabular}{|cc|c|c|}
\hline
\multicolumn{2}{|c|}{Zero-temperature phases}& $\log (\rho_0 / \rho_1)$ & $J$ \\ \hline \hline
\multicolumn{1}{|c|}{\multirow{2}{*}{$\ve_2 < 2 \ve_1$}} & $F < \ve_1$ &
$+\infty$ & $0$ \\ \cline{2-4}
\multicolumn{1}{|c|}{}& $F > \ve_1$ &
$+\infty$ & $+\infty$ \\ \hline
\multicolumn{1}{|c|}{\multirow{3}{*}{$\ve_2 > 2 \ve_1$}} & $F < \ve_1$ &
$+\infty$ & $0$ \\ \cline{2-4}
\multicolumn{1}{|c|}{}& $\ve_1 < F < \ve_2 - \ve_1$ &
$-\infty$ & $0$ \\ \cline{2-4}
\multicolumn{1}{|c|}{}& $\ve_2 - \ve_1 < F $ &
$-\infty$ & $+\infty$ \\ \hline
\end{tabular}
\caption{Zero-temperature phases of the driven three-level model.
In all cases $\log(\rho_2/\rho_{0,1}) \to -\infty$ for $T \to 0^+$.}
\label{table_cases}
\end{table}

\section{Steady heat capacity}\label{sec:heatcapacity}

Now we come to the main topic of this paper, which is a gauge-invariant construction of the heat capacity of driven systems in their steady states. We present here a simplified version of the argument given in \remove[JD]{Ref.}~\cite{new}, which is also close to the original derivation for weakly irreversible heat conducting systems~\cite{cera}.

To keep a system in a nonequilibrium steady state, work needs to be continuously pumped into it and the same amount of energy is then dissipated to the heat bath, cf.~\eqref{dissipation}. When the temperature of the bath changes, the system relaxes to new steady conditions with a generally different rate of dissipation. Therefore, a \emph{direct} calorimetric experiment has to separate the work/heat required to maintain the nonequilibrium conditions from the extra energy needed for the thermodynamic transition to another steady state. It has been proven that this separation can generally be done in a consistent way~\cite{pump,ha} and one obtains an expression for the \emph{excess heat} that is geometric in the quasistatic limit. The latter means that when the temperature changes are made infinitely slow, the corresponding excess heat remains finite and it only depends on the shape of the trajectory in the system-parameter space; i.e., it becomes independent of the speed of the temperature changes. Taking these important results into account, we can actually skip the full derivation and consider just a single elementary step in such a quasistatic process, namely the energy changes along a small and sudden temperature jump.

Let the system be initially prepared at the steady state corresponding to temperature
$T - \de T$ which is suddenly changed to $T$ at time zero:
\begin{equation}
  T(t) =
  \begin{cases}
    T - \de T & \text{ for } t < 0
  \\
    T & \text{ for } t \geq 0
  \end{cases}
\end{equation}
The heat flowing to the bath 
upon
the system's relaxation to the new steady state is obtained \change[JD]{from}{by} summing up the energy quanta exchanged with the bath along all transitions. The expected heat up to a time $t$ along the relaxation process is
\begin{equation}
  Q(t) = \Bigl\langle {\sum}^{(\om)} q_{i \rightarrow j}^b \Bigr\rangle_{\rho^{(T - \de T)}}
\end{equation}
where the sum runs over all transitions occurring along a realization
$\om$\change[JD]{ and t}{. T}he expectation is with respect to the stochastic process at temperature $T$ but started from the initial distribution
$\rho^{(T - \de T)}$, which coincides with the stationary distribution at temperature
$T - \de T$.
By the Markov property of the dynamics, the heat expectation can be written as
\begin{equation}\label{Q-propagator}
  Q(t) = \int_0^t \sum_{i,j} \bar q_j\, P(j,\tau \rel i,0)\,
  \rho^{(T - \de T)}_i\,\id\tau
\end{equation}
Here $P(j,\tau \rel i,0)$ is the transition probability (or propagator) with respect to the dynamics at temperature $T$, and
\begin{equation}\label{heatquanta}
  \bar q_i = \sum_{j,\,b} k_{i \rightarrow j}^b\,
  q_{i \rightarrow j}^b
\end{equation}
is the expected amount of heat per unit of time provided the system is at state $i$. Comparing with~\eqref{dissipation} we check that
$\frq = \langle \bar q \rangle$.
Expanding the initial distribution up to the linear order in $\de T$ and writing the propagator in terms of a generator of the Markov process, \change[JD]{formula}{equation}~\eqref{Q-propagator} gets the form
\begin{equation}\label{fullheat}
  Q(t) = \frq\, t -
  \delta T \int_0^t \sum_i
  \frac{\partial\rho_i}{\partial T}\,(e^{\tau L} \bar q)_i\,\id\tau
\end{equation}
Here
\begin{equation}\label{generator}
  (L g)_i = \sum_{i,\,b} k_{i \rightarrow j}^b (g_j - g_i)
\end{equation}
is the backward Kolmogorov generator, evaluating the expected speed of change of a function $g$ provided the system is at state $i$. Clearly, the first term in~\eqref{fullheat} is  the steady dissipation up to time $t$, whereas the second term is the extra contribution which accounts for the thermodynamic process connecting the different steady states. Its limiting value
\begin{equation}
  \de Q^\text{ex} = \lim_{t \to \infty} [Q(t) - \frq\,t]
\end{equation}
is the announced excess heat for the case of our elementary heat process. It can be written in the form
\begin{equation}\label{excess}
  \de Q^\text{ex} = -\de T \sum_i \frac{\partial\rho_i}{\partial T}\,V_i
\end{equation}
where we have introduced the \emph{quasipotential}
\begin{equation}\label{quasipotential}
  V_i = \int_0^{+\infty} [(e^{\tau L} \bar q)_i - \frq]\,\id\tau
\end{equation}
One checks that the integral converges whenever the dynamics has a spectral gap. Note that for discrete processes with \change[JD]{finitely many}{a finite number of} states this boils down to the condition that the model has a unique stationary distribution.
By definition, the quasipotential is gauge-invariant and centralized around zero
$\langle V \rangle = 0$.

In analogy with equilibrium thermodynamics, the steady heat capacity is naturally defined
to quantify \add[JD]{the} amount of \remove[JD]{the} heat (excess) as
$-\de Q^\text{ex} = C_\text{neq} \de T$, where the minus sign comes from our convention that positive heat flows \emph{out} of the system.
In terms of the quasipotential, the steady heat capacity equals
\begin{equation}\label{main}
  C_\text{neq} = \sum_i \frac{\partial\rho_i}{\partial T}\,V_i
  = - \Bigl\langle \frac{\partial V}{\partial T} \Bigr\rangle
\end{equation}
Before turning to explicit calculations on our test examples, we first look into general properties of the quasipotential to better understand in what sense it can be seen as a generalized energy function. We also indicate how it \change[JD]{enters the quasistatic energetics in a more general context}{plays a more general role in the quasistatic energetics}.

\subsection{Properties of the quasipotential}\label{sec:quasipotential}

We start by writing another expression for the quasipotential. Applying the generator $L$ to both sides of~\eqref{quasipotential} and using \add[JD]{the fact} that $L\, [\text{const}] = 0$, we get the relation
\begin{equation}\label{quasipotential-equation}
  (L V)_i = \frq - \bar q_i
\end{equation}
which specifies $V$ up to a constant, the latter being fixed by the condition
$\langle V \rangle = 0$.
In this way, the steady heat capacity~\eqref{main} is most conveniently evaluated by solving the (in our case algebraic) equation~\eqref{quasipotential-equation} with the linear operator $L$ given by~\eqref{generator}; the stationary quantities
$\rho_i$ and $\frq$ can be found from~\eqref{stationary} and~\eqref{dissipation}.\\

To first check the consistency with equilibrium thermodynamics, let the heat quanta
$q_{i \rightarrow j}^b$ be given as in~\eqref{balance-micro}, via energy levels
$\ve_i$ and work functions $w_{i \rightarrow j}^b$. Assuming that the latter derive from a potential, they can be transformed out by defining new energy levels $\hat\ve_i$, \cf~\eqref{gauge}. This system obeys detailed balance with respect to the canonical stationary distribution $\rho_i \propto \exp(-\hat\ve_i / T)$ and the steady dissipation vanishes, $\frq = 0$. In this case the heat function~\eqref{heatquanta} reads
$\bar q_i = \sum_{j,\,b} k_{i \rightarrow j}^b (\hat\ve_i - \hat\ve_j)$ and the quasipotential becomes $V_i = \hat\ve_i - \langle \hat\ve \rangle$, i.e., it essentially coincides with the energy levels up to a (temperature-dependent) constant ensuring the proper normalization. As a result we recover the familiar expression for the equilibrium heat capacity in terms of the temperature-energy response:
$C_\text{eq} = \partial \langle \hat\ve \rangle / \partial T$. Note that although the quasipotential generally depends on both the state $i$ and the temperature $T$, these variables are decoupled in equilibrium, in the sense that the differences $V_i - V_j$ are temperature-independent and the derivatives $\partial V_i / \partial T$ no longer depend on $i$.
Out of equilibrium the temperature--state coupling in the quasipotential $V$ remains nontrivial\change[JD]{;}{:} in particular, the quasipotential differences
$V_i - V_j$ still depend on temperature.

As an alternative to the gauge-invariant form~\eqref{main}, we can choose a specific gauge with some energy levels $\ve_i$ and work functions
$w_{i \rightarrow j}^b$. Then the quasipotential reads
$V_i = \ve_i - \langle \ve \rangle + \breve V_i$ with
$(L \breve V)_i = \frq - \bar w_i$,
$\bar w_i = \sum_{j,\,b} k_{i \rightarrow j}^b\,w_{i \rightarrow j}^b$.
So the heat capacity obtains the form
\begin{equation}\label{capacity-gauge}
  C_\text{neq} = \frac{\partial \langle \ve \rangle}{\partial T} -
  \Bigl\langle \frac{\partial \breve V}{\partial T} \Bigr\rangle\
\end{equation}
Obviously, in this representation each of the right-hand side terms depends on the gauge. Whenever the latter can be fixed so that $\breve V$ becomes small enough, making the second term in~\eqref{capacity-gauge} negligible with respect to the first term, we obtain an approximative equality between the heat capacity and the temperature-energy response.
Note, however, that the stationary distribution is
in general not Boltzmannian with respect to the energy levels $\ve_i$ fixed by the gauge. A \change[JD]{most simple}{basic} example where such a simplification occurs is the close-to-equilibrium regime\add[JD]{,} with the gauge being naturally fixed by \remove[JD]{a} reference equilibrium dynamics. Other examples are provided by various models in either high-temperature or low-temperature regimes, even without the restriction to weak nonequilibrium; see section~\ref{sec:results}.

The last argument also indicates that in specific models and regimes, the \emph{a priori} ambiguous notion of energy levels for open systems driven by non-potential forces can be given a more precise meaning by using the gauge invariance to fix the gauge in \change[JD]{some most suitable way}{an appropriate manner}. We briefly touch \add[JD]{on} this issue in the following section.

\subsection{General quasistatic processes}

Although the above construction of the excess heat was specified for \change[JD]{the}{a} thermodynamic process in which only the temperature changes, it can obviously be generalized to arbitrary quasistatic processes. Here the word `quasistatic' refers to slow time-dependent transformations under which the system evolution can be seen as a sequence of small sudden changes\add[JD]{,} as above\change[JD]{, each of them leaving enough time}{. There is sufficient time between these changes} for the system to relax to new steady conditions. Upon this generalization, the heat excess takes the form, cf.~\eqref{excess},
\begin{equation}\label{general}
  \de Q^\text{ex} = -\sum_i V_i\, \de\rho_i =
  \langle \de V \rangle
\end{equation}
Within a fixed gauge we have
$V_i = \ve_i - \langle \ve \rangle + \breve V_i$ as above, yielding the
quasistatic energy balance relation
\begin{equation}\label{general-gauge}
  \de \langle \ve \rangle =  - \de Q^\text{ex} + \langle \de \ve \rangle + \langle \de \breve V \rangle
\end{equation}
which represents the change of mean energy as the sum of the quasistatic excess heat, the quasistatic work of potential forces and the quasistatic excess work of remaining non-potential forces; the last term being absent in equilibrium thermodynamics.

Again, whenever the quasipotential $V_i(T)$ is such that its variables $i$ and $T$ approximatively decouple\add[JD]{,} then it can always be written in the form
$V_i \simeq \ve_i + \text{const}(T)$ with some temperature-independent energy levels $\ve_i$; then $\text{const}(T) = -\langle \ve \rangle$ and hence
$\breve V \simeq 0$.
In such a case the energy levels can be considered \change{as}{to be} well-defined, in spite of the presence of (possibly strong) non-potential forces. Note however that even in such a case the nonequilibrium nature of the dynamics 
is still made apparent by
the nontrivial relation between the energy levels $\ve_i$ and the stationary distribution $\rho$.

\section{Results for model systems}\label{sec:results}

Now we \change[JD]{specify}{apply} the general considerations of the previous section to the two models introduced in \change[JD]{the beginning}{section II}.

\subsection{\change[JD]{m}{M}odel I}\label{subsec:I-results}

\begin{figure}[t]
\includegraphics[width=8cm]{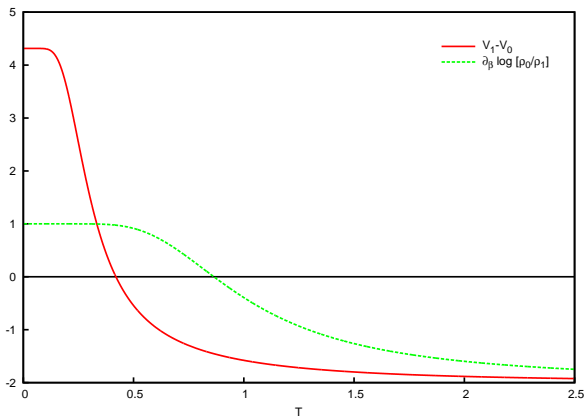}\\
  \caption{The quasipotential gap $\De V = V_1 - V_0$ compared with the gap
  $G = \partial_\be \log\,[\rho_0 / \rho_1]$. The model parameters are $U = 1$ and
  $\Phi = 3$.}\label{2level-gap}
\end{figure}
First we consider the two-level model with asymmetric channels which was introduced in \change[JD]{Sec.}{section}~\ref{sec:2-level}. In terms of the \emph{a priori} gauge with the energy levels
$\ve_0 = 0$, $\ve_1 = U$ and the work functions
$w_\uparrow^\pm = \pm F$ and $w_\downarrow^\pm = \mp F$, the quasipotential is given as
$V_i = \ve_i - \langle \ve \rangle + \breve V_i$\add[JD]{,} with $(L \breve V)_i = \frq - \bar w_i$ and
$\bar w_0 = F(k_\uparrow^+ - k_\uparrow^-)$,
$\bar w_1 = F(k_\downarrow^- - k_\downarrow^+)$;
\cf \change[JD]{Sec.}{section}~\ref{sec:quasipotential}. In fact, we only need to determine the gap in the quasipotential,
$\De V = U + \breve V_1 - \breve V_0$, where the latter difference is obtained from
$(L \breve V)_1 - (L \breve V)_0 = \bar w_0 - \bar w_1$. As a result,
\begin{equation}
  \De V = U + F \frac{\tanh\bigl(\frac{U}{2T}\bigr) \tanh\bigl(\frac{F}{2T}\bigr)
  - \tanh\bigl(\frac{\Phi}{2}\bigr)}{1 - \tanh\bigl(\frac{U}{2T}\bigr)
  \tanh\bigl(\frac{F}{2T}\bigr) \tanh\bigl(\frac{\Phi}{2}\bigr)}
\end{equation}
The steady heat capacity is then obtained from \change[JD]{eq.}{equation}~\eqref{capacity-gauge}\add[JD]{,} which for the two-level model simplifies to
\begin{equation}\label{capacity-explicit}
  C_\text{neq} = \frac{\rho_0 \rho_1}{T^2}\, G\, \De V
\end{equation}
with the shorthand $G = \partial_\be \log (\rho_0 / \rho_1)$, $\be = 1/T$.

\begin{figure}[t]
\includegraphics[width=8cm]{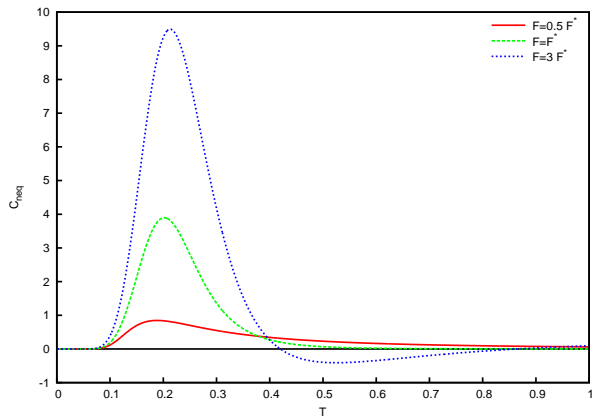}\\
  \caption{The temperature dependence of the heat capacity for subcritical, critical and supercritical driving. The model parameters are $U = 1$ and $\Phi = 3$.}\label{2level-CT}
\end{figure}

Note first that for $F = 0$ one gets $\De V = G = U$ and we obtain a well-known formula for the heat capacity of an equilibrium two-state model. Away from equilibrium the picture becomes far more complicated since both energy-dimensional quantities
$G$ and $\De V$ are now different and generally not related in a simple way. Moreover, they can obtain opposite signs for large enough driving forces, which then results in negative values of the steady heat capacity, see Figs.~\ref{2level-gap}--\ref{2level-CT}. Next we separately analyze three asymptotic regimes.\\

\emph{High temperatures.}
For large temperature\change[JD]{s}{ values} the quasipotential gap is
$\De V = U - F \tanh(\Phi / 2) + O(1/T^2)$, which coincides with the asymptotics of $G$ as obtained from \change[JD]{eq.}{equation}~\eqref{2level-rho-highT}. Hence the heat capacity equals
\begin{equation}\label{2-level-capacity}
  C_\text{neq} = \frac{\bigl[
  U - F \tanh \bigl( \frac{\Phi}{2} \bigr) \bigr]^2}{4 T^2} +
  o\Bigl( \frac{1}{T^2} \Bigr)
\end{equation}
We see that
$F^* := U / \tanh(\Phi / 2)$ is a critical value of the driving\add[JD]{,} above which \remove[JD]{both} the system exhibits a population inversion and \add[JD]{also} the gap $\De V$ changes sign. As a result, for any $F \neq F^*$ the heat capacity is asymptotically strictly positive and decaying as $1 / T^2$, i.e., similarly as in equilibrium. Note that the asymptotic equality
$\De V \simeq G$ remains true even for \add[JD]{a} driving force $F$ much large\add[JD]{r} than the model parameter $U$, the original meaning of which as an energy gap \change[JD]{becomes then}{then becomes} meaningless. Instead, there is another gauge that becomes natural here: \change[JD]{B}{b}y making the transformation~\eqref{gauge} with $\psi_0 = 0$ and $\psi_1 = -F \tanh(\Phi / 2)$, we obtain ``renormalized'' energy levels with the gap
$\hat\ve_1 - \hat\ve_0 = U - F \tanh(\Phi / 2)$, which is directly seen in the leading asymptotics $\De V \simeq G \simeq \hat\ve_1 - \hat\ve_0$. After this transformation, the residual non-potential forces contribute to the heat capacity only by correction
$o(1 / T^2)$. In this sense the full high-temperature regime away from the critical value $F^*$ is to be understood as essentially close to equilibrium, but with the renormalized energy levels $\hat\ve_i$ and the corresponding Boltzmann stationary distribution. From this point of view the observed population inversion at high temperatures and strong driving is only an artifact of describing the model in terms of ``unphysical'' energy levels $\ve_i$.\\

\begin{figure}[t]
\includegraphics[width=8cm]{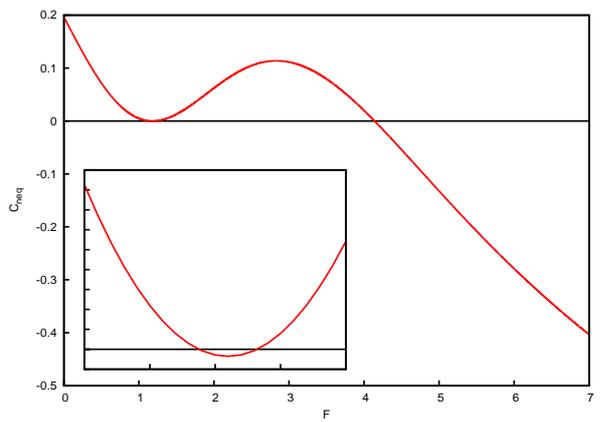}\\
\caption{Steady heat capacity as a function of the driving, with the parameters
  $U = 1$, $\Phi = 3$, and $T = 1$. The tiny region of negative heat capacity in the vicinity of the critical driving is zoomed in.}
\label{2level-CF}
\end{figure}

\emph{High temperatures --- critical.}
We have seen that the value $F = F^*$ plays a special role since \change[JD]{there}{in this case} the above gauge transformation leads to degenerate energy levels\add[JD]{,} and therefore the heat capacity becomes zero up to order $1 / T^2$. More detailed calculations reveal
that both gaps $\De V$ and $G$ are of order $1/T^2$ which yields an anomalously fast-decaying heat capacity,
\begin{equation}
  C_\text{neq} = \frac{U^6}{64 T^6 \sinh^4 \bigl( \frac{\Phi}{2} \bigr)}
  + o(1/T^6)
\end{equation}
The existence of the high-temperature critical driving leads to the following subtle phenomenon: \change[JD]{T}{t}here is a temperature curve $T = T^1(F)$ along which
$\De V = 0$ and another one, $T = T^2(F) > T^1(F)$, on which $G = 0$. Both curves have the identical leading asymptotics, for $F > F^*$,
\begin{equation}
  \frac{1}{T^{1,2}(F)} = 2 \sinh \bigl( \frac{\Phi}{2} \bigr) \sqrt{\frac{2(F - F^*)}{U^3 \sinh \Phi}} + o((F - F^*)^{1/2})
\end{equation}
On both curves the heat capacity vanishes and they form the boundary of a tiny region in the $(T,F)-$space inside which $C_\text{neq}$ exhibits negative values.
Its full dependence on the driving for a fixed intermediate temperature is depicted on Fig.~\ref{2level-CF} where the above mentioned region has been zoomed in. Notice the negative values of the heat capacity for large $F$; this is a strong-nonequilibrium effect and we may expect that no gauge transformation would significantly simplify the thermodynamic description in this region due to a strong temperature-dependence of the quasipotential gap $\De V$.\\

\emph{Low temperatures.}
The quasipotential gap $\De V$ has the low-temperature asymptotics
\begin{equation}\label{gap-lowT}
  \De V = U + |F| + O(|F|\,e^{- \min\{U,|F|\} / T})
\end{equation}
in which the temperature dependence emerges only in the exponentially small correction (along the limit $T \to 0^+$). This suggest\add[JD]{s that it is appropriate} to make the gauge transformation with
$\psi_0 = 0$ and $\psi_1 = F$\add[JD]{,} to define the ``renormalized'' energy levels
$\hat\ve_0 = 0$ and $\hat\ve_1 = U + |F|$ through which the heat capacity gets the simplified approximate form $C_\text{neq} \simeq \partial\langle \hat\ve \rangle / \partial T$, i.e, with only a negligible contribution from the second term in~\eqref{capacity-gauge}. One checks by comparing with the exact result that this intuition is indeed correct.
The apparent disagreement between the low-temperature asymptotics of the gaps
$\De V$ and $G$, cf.~\eqref{2-occupation} and \eqref{gap-lowT}, indicates that the low-temperature regime corresponds to strong nonequilibrium with non-Boltzmannian statistics. Formally, it can be described by an effective temperature defined by
$\log (\rho_1 / \rho_0) = -(\hat\ve_1 - \hat \ve_0) / T^\text{eff}$, explicitly $T^\text{eff} = T (1 + |F| / U) > T$.
Using \add[JD]{the fact} that $C_\text{neq} \simeq (1 + |F|/U)\,\partial \langle \hat\ve \rangle / \partial T^\text{eff}$, we can trace back the exponential decay of the heat capacity for
$T \to 0^+$ to the exponential suppression of thermal excitation, \change[JD]{analogously as in}{which is analogous to} the equilibrium Third law.
\begin{figure}[t]
\includegraphics[width=8cm]{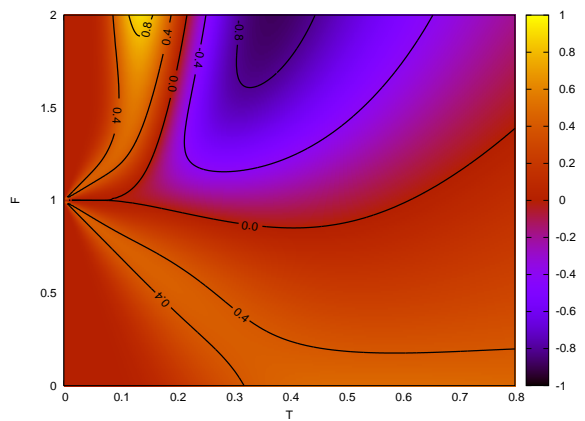}\\
 \caption{Curves of constant heat capacity in the $(T,F)-$plane for the three-level model. The parameters are $\ve_1 = 1$, $\ve_2 = 3$, and $A_{0,1} = 1$, $A_{1,2} = 2$, $A_{2,0} = 4$.}\label{color}
\end{figure}

Since $\De V \simeq U + |F| > 0$\add[JD]{,} and
recalling that our model exhibits no population inversion in the zero-temperature limit, we conclude that the steady heat capacity remains strictly positive at low temperatures. In particular, it does not exhibit any transition at $F = U$ where the system undertakes a change between the ``insulator'' and the ``conductive'' transport regime\add[JD]{s}.\\

We finish this section by indicating how to extend the above approximate description of the low-temperature behavior to arbitrary temperatures and driving forces. \emph{Formally} defining the effective temperature
$T^\text{eff} = \De V / \log(\rho_0 / \rho_1)$, we can write the heat excess~\eqref{general} in the form of a Clausius equality
\begin{equation}
  -\de Q^\text{ex} = T^\text{eff} \de S\,,\quad
  S = -\sum_{i=1,2} \rho_i \log \rho_i
\end{equation}
with $S$ the Shannon entropy of the stationary distribution $\rho$.
In this framework the heat capacity obtains the form
$C_\text{neq} = T^\text{eff} \de S / \de T$.
In contrast with the above low-temperature regime, the effective temperature now becomes a nontrivial function of $T$; for example, it becomes zero on the critical line
$T = T^1(F)$.
Obviously, such a representation in terms of a (single) effective temperature has no straightforward extension to models with a larger number of states\add[JD]{,} and \change[JD]{serious}{significant} modifications are needed. Some extensions of the Clausius relation to nonequilibrium and its limitations have been studied in~\cite{ha,sh}.

\subsection{Model II}

\begin{figure*}
\includegraphics[width=.9\textwidth]{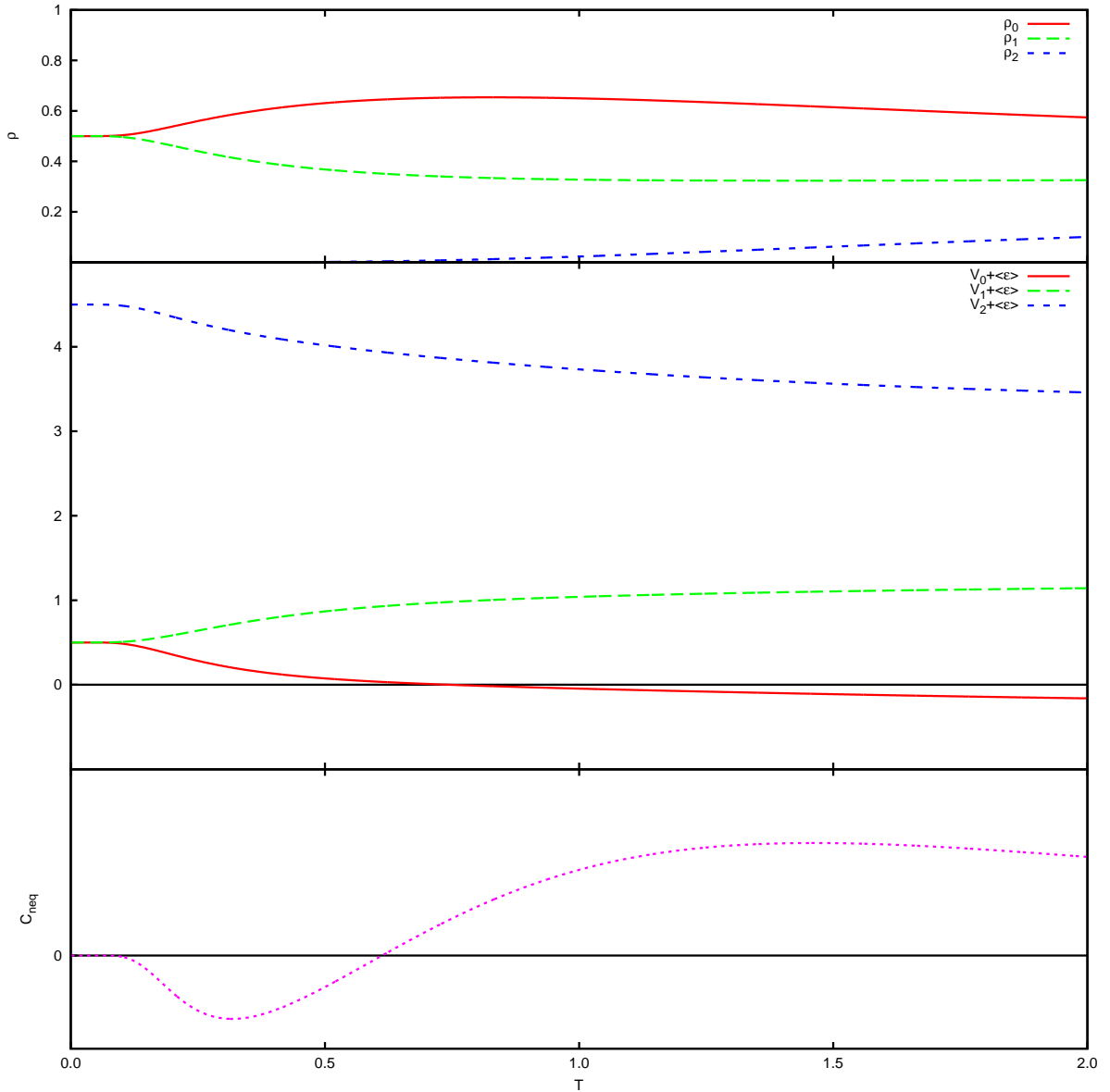}\\
\caption{Steady heat capacity of the three-level model for the critical driving force
  $F = \ve_1 = 1$; the model parameters being the same as in Fig.~\ref{color}. The upper, the middle and the lower parts subsequently display the temperature dependence of the stationary occupations, the quasipotential (shifted by the mean energy $\langle \ve \rangle$), and the steady heat capacity.}
\label{3level-crit}
\end{figure*}
For the three-level model we skip the detailed analysis and \remove[JD]{we} only concentrate on some new features that were not seen in the previous two-level example. Therefore we consider the case $\ve_2 > 2\ve_1$ in which the system exhibits a zero-temperature transition in the stationary occupations, cf.\change[JD]{~Sec.}{section}~\ref{sec:3-levels-intro}.
The heat capacity in the
$(T,F)-$plane is depicted in Fig.~\ref{color} where we see that $(T=0,F=\ve_1)$ is an accumulation point of the curves of constant heat capacity.

In order to better understand the behavior we look into the critical case $F = \ve_1$ in more detail\change[JD]{;}{, and} the results are summarized in Fig.~\ref{3level-crit}.
At zero temperature the states `$0$' and `$1$' degenerate into a single energy level, both in the sense of stationary occupations, $\rho_0 = \rho_1 = 1/2$, \change[JD]{as well as}{and} in the sense of the quasipotential, $V_0 = V_1$. Increasing the temperature, the occupation of state `$0$' also increases (\add[JD]{in} this way behaving like an excited state), whereas the quasipotential satisfies
$V_1 > V_0$, meaning that the degeneracy gets removed and a positive energy gap opens between states `$0$' (lower) and `$1$' (higher). As a direct consequence of these opposite tendencies the steady heat capacity becomes negative at $F = \ve_1$ and low temperatures, with an anomalously fast decay to zero for $T \to 0^+$ due to the zero-temperature degeneracy of both states. Note that although the presence of state `$3$' is essential for breaking the detailed balance and for the nonequilibrium features of our model, it does not directly enter low-temperature energetics\change[JD]{ and it}{. It also} does not substantially contribute to the heat capacity until high enough temperatures where its occupation becomes relevant.

\section{Conclusions and open problems}\label{sec:conclusion}

In this paper we have studied, via simple examples, the theoretical proposal that nonequilibrium steady states can be assigned a generalized heat capacity characterizing the amount of heat absorbed along quasistatic changes of the environment temperature. An essential point is that here we are concerned with contributions to the total dissipated heat which are intrinsically due to the (slow) transformations, whereas the dominant contribution always comes from the omnipresent steady dissipation. Those ``extra costs'' needed to (slowly) drive the system to a nearby steady state can be measured in terms of the excess heat which is by now a well-defined theoretical concept.
The resulting heat capacity as defined upon the excess heat generally differs from the temperature-energy response of the system and, in fact, it is independent of the way the system's energy is defined. This is particularly important for systems maintained out of thermal equilibrium by non-conservative forces for which the energy function depends on the gauge chosen.

As a particular example, we have studied in detail the two-level model driven by non-potential forces acting along two distinct transition channels. It has been demonstrated that its steady heat capacity can obtain negative values when far from thermal equilibrium. This phenomenon was traced back to a large discrepancy between two characteristic energy quantities which appear in the dynamical problem and which coincide under detailed balance condition: (1) the spacing between ``renormalized'' energy levels as defined via the quasistatic excess heat along relaxation processes started from different states, and (2) an effective energy gap associated with the stationary occupation statistics. We have also proposed that the gauge invariance corresponding to different choices of both the energy levels and the non-potential work functions may sometimes be exploited to simplify the quasistatic energetics\add[JD]{,} in the sense that the work of (gauged) non-potential forces is made irrelevant. In such a case the heat capacity is approximatively given in terms of the temperature-energy response, analogously to equilibrium. It was also proposed that the quasistatic excess heat for driven two-level systems can be expressed via a Clausius-type relation with the usual Shannon entropy and an appropriate effective temperature. We have further discussed properties of a driven three-level model in a regime where the system exhibits a population inversion between the two lower levels and negative heat capacity at arbitrarily low temperatures.

The most relevant open problem seems to be the experimental accessibility of the quasistatic excess heat, in particular far from thermal equilibrium where the heat capacity has been shown to exhibit qualitatively new features. We leave the analysis of experimentally more relevant 
mesoscopic
systems to a future work. Some deeper theoretical insight\add[JD]{s} into a general thermodynamic role \change[JD]{of}{for} the nonequilibrium quasipotential as well as into its relation to the stationary occupations are needed. This problem also appears to fit the recent discussions on the status and possible extensions of the Clausius relation away from thermal equilibrium~\cite{ha,sh}.

The present work also suggests that even far from equilibrium and possibly beyond the scope of generalized Clausius relations, there still may be another simplification in some cases due to the existence of a preferred gauge in which the energy levels become essentially temperature-independent. As it was demonstrated on the low-temperature regime of our two-level model with distinctly non-Boltzmannian statistics (under the original temperature), this proposal indeed goes beyond usual close-to-equilibrium considerations. It remains to be understood better whether such an essential removal of non-conservative contributions to the quasistatic energetics may lead to useful relations between quantities characterizing far-from-equilibrium steady states. Other open questions include the relation of the steady heat capacity to nonequilibrium fluctuations, the understanding of its low-temperature patterns or derivation of general lower bounds on the (possibly negative) heat capacity.

\begin{acknowledgements}
We are grateful to Christian Maes and Tom\'a\v{s} Novotn\'y for many useful discussions.
J.P.\ benefits from the Grant no.~51410 (the Grant Agency of Charles University) and from the project SVV-265301 (Charles University). K.N.\ acknowledges  the  support  from  the  Academy of Sciences of the Czech Republic under Project No.~AV0Z10100520. E.B.~acknowledges financial support from the FWO project G.0422.09N.
\end{acknowledgements}



\begin{thebibliography}{10}

\bibitem{sta}
N.~O.~Birge and S.~R.~Nagel,
Phys.~Rev.~ Lett.~54, 2674 (1985)

\bibitem{cera}
J.~Cerro and S.~Ramos,
Ferroelectrics Lett.~16, 119 (1993)

\bibitem{nd}
J.~K.~Nielsen and J.~C.~Dyre,
Phys.~Rev.~B 54, 15754 (1996)

\bibitem{new}
E.~Boksenbojm, C.~Maes, K.~Neto\v{c}n\'{y}, and J.~Pe\v{s}ek,
Europhys.~Lett.~96, 40001 (2011)

\bibitem{pump}
N.~A.~Sinitsyn and I.~Nemenman,
Europhys.~Lett.~77, 58001 (2007)

\bibitem{ha}
T.~S.~Komatsu, N.~Nakagawa, S.~Sasa, and H.~Tasaki,
Phys.~Rev.~Lett.~100, 230602 (2008);
J.~Stat.~Phys.~134, 401 (2009)

\bibitem{seifert}
U.~Seifert,
In: J.~K.~G.~Dhont, G.~Gompper, G.~Nagele,~D. Richter, and R.~G.~Winkler (eds.), 
Soft Matter. From Synthetic to Biological Materials: Lecture Notes of the 39th Spring School 2008 (Institute of Solid State Research, Forschungszentrum Julich, 2008) B.5

\bibitem{LS}
J.~L.~Lebowitz and H.~Spohn,
J.~Stat.~Phys.~95, 333 (1999)

\bibitem{sh}
T.~Sagawa and H.~Hayakawa,
arXiv:1109.0796v1 [cond-mat.stat-mech]

\end{thebibliography}
\end{document}